\documentclass[namedreferences]{solarphysics}
\usepackage[optionalrh]{spr-sola-addons} 
\usepackage{graphicx}        
\usepackage{color}           
\usepackage{url}             

\newcommand{\Rs}{$ R_{\odot}$}

\newcommand{\aap}{    {\it Astron. Astrophys.}}

\newcommand{\apj}{    {\it Astrophys. J.}}

\newcommand{\nat}{    {\it Nature}}

\newcommand{\solphys}{{\it Solar Phys.}}

\begin{document}

\begin{article}

\begin{opening}

\title{Multi-scale Gaussian normalization for solar image processing}

\author{Huw~\surname{Morgan}$^{1}$\sep
		Miloslav~\surname{Druckm\"uller}$^2$        }
\runningauthor{Morgan \& Druckm\"uller}
\runningtitle{MGN image processing}

   \institute{$^{1}$ Sefydliad Mathemateg a Ffiseg, Prifysgol Aberystwyth, Ceredigion, Cymru, SY23 3BZ
   $^{2}$ Faculty of Mechanical Engineering, Brno University of Technology, 616 69 Brno, Czech Republic\\
                     email: \url{hmorgan@aber.ac.uk} \\ 
             }

\begin{abstract}
Extreme UltraViolet images of the corona contain information over a large range of spatial scales, and different structures such as active regions, quiet Sun and filament channels contain information at very different brightness regimes. Processing of these images is important to reveal information, often hidden within the data, without introducing artifacts or bias. It is also important that any process be computationally efficient, particularly given the fine spatial and temporal resolution of \textit{Atmospheric Imaging Assembly} on the \textit{Solar Dynamics Observatory}  (AIA/SDO) , and consideration of future higher-resolution observations. A very efficient process is described here which is based on localized normalizing of the data at many different spatial scales. The method reveals information at the finest scales, whilst maintaining enough of the larger-scale information to provide context. It also intrinsically flattens noisy regions and can reveal structure in off-limb regions out to the edge of the field of view. The method is also successfully applied to a white light coronagraph observation.
\end{abstract}
\end{opening}

\section{Introduction}

Extreme UltraViolet (EUV) observations currently provide the most important source of information on the low solar corona. As new EUV instruments are developed, the temporal, spatial and spectral resolution becomes ever finer, giving new insight on the coronal and chromospheric structure and dynamics. The \textit{Atmospheric Imaging Assembly} (AIA: \opencite{lemen2011}) aboard the \textit{Solar Dynamics Observatory} (SDO: \opencite{pesnell2012}) provides very fine temporal and spatial resolution of the Sun at multiple wavelengths, and is having a large impact on the field. Even as the community develop methods to digest the huge volume of data from AIA/SDO, new instruments are planned and tested with even finer resolution (e.g. \textit{High-Resolution Coronal Imager} (Hi-C), see \opencite{cirtain2013}). 

Despite the development of automated detection tools for EUV observations (e.g. \opencite{martens2012}), most scientific works begin from visual inspection of images. In particular, the volume of data provided by AIA/SDO is so high that low-resolution images are often used as a starting point to find features of interest. The higher-resolution data are then used for further analysis. Image processing is therefore an important step in analysing the data. The scientific return can be improved by the application of processing which better reveals features in the data, particularly in the early stages of analysis where visual inspection is most important. 

A common approach to process EUV images is simply to display the square root (or a gamma curve transformation), or alternatively the logarithm, of the original pixel values. This is a quick and easy way of reducing the dominance of the image contrast range by a few small bright regions. To reveal dynamic features, time-differencing is used, where a previous image is subtracted from the current image. Such simple processes are commonly used not due to the quality of the output, but due to their simplicity. More advanced image processing methods are not so commonly used due to their complexity and computational expense. 

Wavelet-based techniques have been used by \inlinecite{stenborg2003} and \inlinecite{stenborg2008} to greatly improve the visual information available from the \textit{Extreme Ultraviolet Imaging Telescope} (EIT: \opencite{Delaboudiniere1995}) aboard the \textit{Solar and Heliospheric Observatory} (SOHO) and the \textit{Extreme UltraViolet Imagers} (EUVI: \opencite{howard2002}) aboard the \textit{Solar Terrestial Relations Observatory} (STEREO: \opencite{kaiser2005})  . The technique involves the decomposition of images into different spatial scales, and the filtering/enhancements of features at the multiple resolutions. The technique is computationally expensive yet gives very good results. A less sophisticated yet very efficient technique to reveal features above the limb is based on techniques originally developed for coronagraphs \cite{morgan2006,druckmullerova2011}. Most relevant to this work is the recently-developed Noise Adaptive Fuzzy Equalization (NAFE) method \cite{druckmuller2013}. The method is inspired by adaptive histogram equalization, where local statistics govern the output value of a pixel. The method uses a fuzzy membership function of a Gaussian-weighted local set to enhance structural detail, preserve contextual detail, and to reduce noise. The NAFE method results in very clear images without artifacts and with excellent noise reduction. Its one downside is computational expense.

This work summarizes the problems in visualising features in EUV images (Section \ref{observation}), introduces the new method (section \ref{method}), applies the method to several images from various instruments (section \ref{results}) and closes with a brief summary (section \ref{summary}).

\section{Observations} 
\label{observation}
\begin{figure}    
   \centerline{\includegraphics[width=1.0\textwidth,clip=]{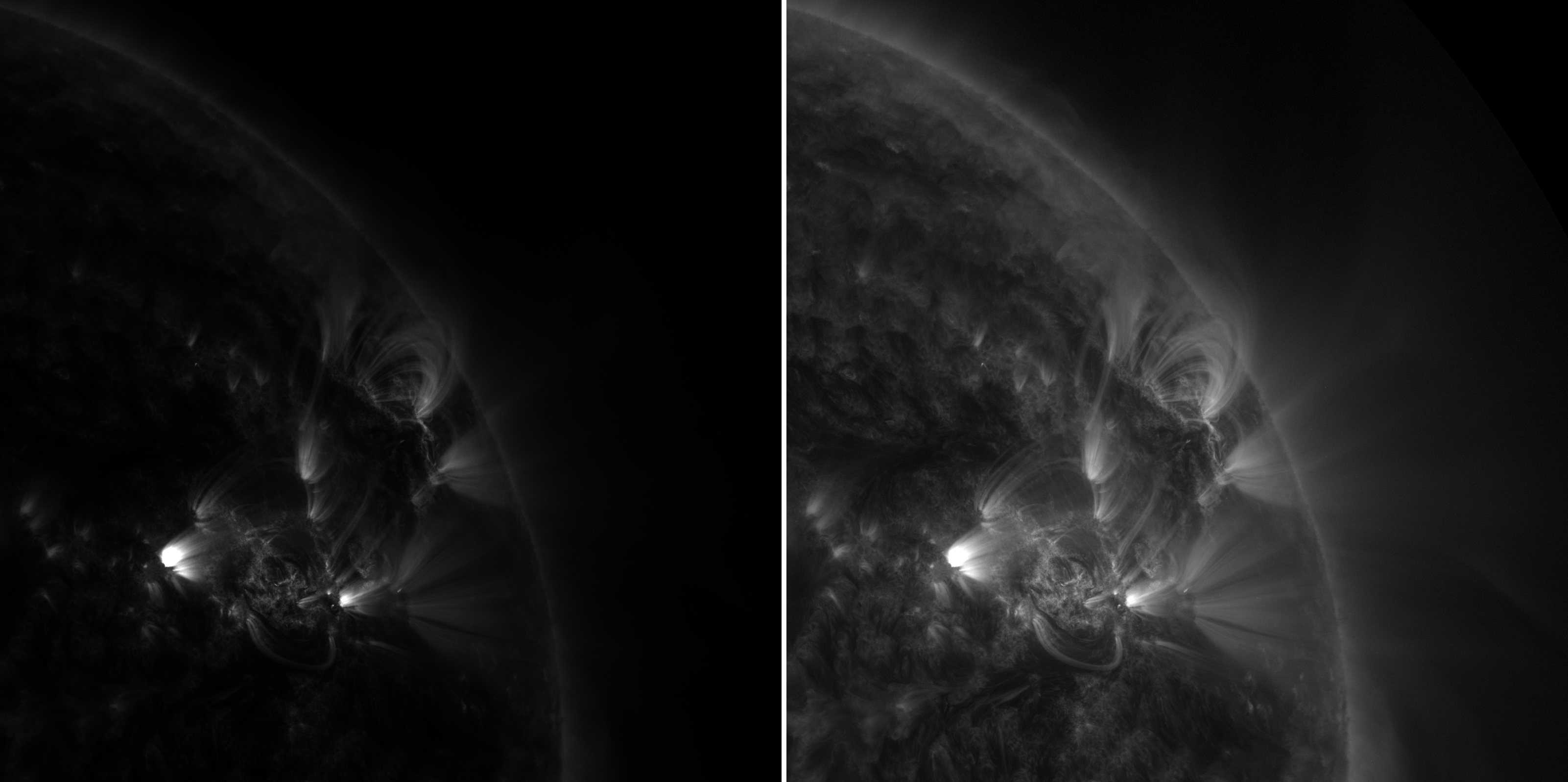}}
   \caption{Left - unprocessed AIA 171 \AA\ image taken at 04 May 2005 00:00 UT showing the north-west disk and off-limb region. Right - the same image with pixel values transformed by taking the square root.}
    \label{fig1}
  \end{figure}

An observation from 04 May 2005 00:00 UT by the 171 \AA\ channel of AIA is used as a working example throughout this section. The data is first pre-processed using the standard SDO Solarsoft software `aia\_prep'. This observation is shown without further image processing in the left panel of Figure \ref{fig1}. The 171 \AA\ channel of AIA has a narrow bandwidth which is dominated by emission from highly-ionised iron which has a formation temperature which peaks at $\sim$0.8 MK. The intensity of this image is therefore a function of the emitting plasma temperature and density. The line-of-sight and optical thickness of the plasma also has an effect on the measured intensity, as well as small contributions from other weaker spectral lines which share the same bandpass. The left panel of Figure \ref{fig1} illustrates the main challenge of revealing information in EUV images. For various physical reasons (mostly due to density), the display range is dominated by the large difference between dark regions and the small bright regions near, or at, the base of active regions. Most of the structural detail appears dark and is hidden. A quick and easy way of revealing some of this structure is by taking the square root of the image (processing of this type is generally known as a gamma transformation). This is shown in the right panel of Figure \ref{fig1}. Although a large improvement on the unprocessed image, there is still much hidden structure, and bright areas become `washed out'. In particular, there is very little visible structure off the limb. The same is true of applying a logarithmic transformation, or any one-to-one mapping between input and output.

\begin{figure}    
   \centerline{\includegraphics[width=1.0\textwidth,clip=]{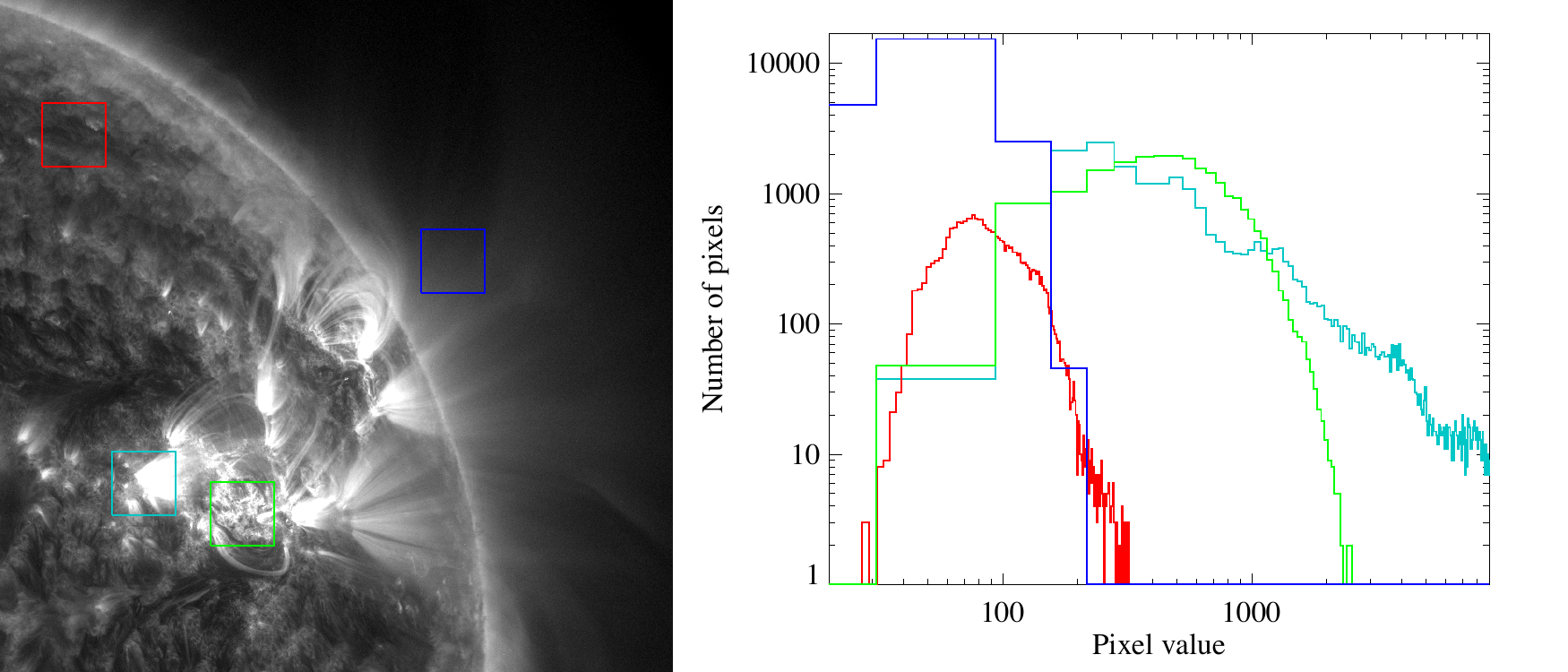}}
   \caption{Left - four square regions are bounded and coloured: red - quiet Sun, cyan - active region base, green - active region and blue - off-limb. Right - histograms of values within each square box.}
    \label{fig2}
  \end{figure}

The dominance of the image contrast range by small bright regions can be shown quantitatively by selecting regions and plotting histograms of pixel values. This is shown in Figure \ref{fig2} for four areas - quiet Sun, active region base, active region, and off-limb. Note that these are pixel values normalized by the exposure time. The top $\sim$75\% of brightness values is taken by the small very brightest region of the active region (cyan colour). This leaves the information from all other regions in the lower 25\%. In fact, the off-limb and quiet Sun regions have pixel values restricted to below $\sim$300. Furthermore, the bright active regions contain a large number of pixels which share the low values of the quiet Sun. There are sharp boundaries between very dark and very bright regions. These characteristics become more extreme when flares occur.

Any image processing which attempts to reveal the information hidden in EUV images must deal with the large range of brightness values, preserve the large spatial gradients in brightness and respect the fact that even the brightest regions contain dark patches. Another challenge is the existence of structures at many spatial scales. We cannot apply edge-detection or enhancement exclusively at small scales since this will enhance some features whilst other structures and the larger-scale context is lost. Furthermore, with full resolution AIA images of $4096\times4096$ pixels taken every 11s (and the anticipation of the resolution of future instruments), computational efficiency is very important.

\section{Method - Multiscale Gaussian Normalization (MGN)}
\label{method}

Barring spurious calibration errors, the brightness values in the original EUV image will be positive. Let $B$ be the original image normalised by exposure time. $k_w$ is a 2D Gaussian kernel of width $w$ pixels in the $x$ and $y$ image dimensions. Throughout this work, $w$ denotes the one-sigma width of the Gaussian. A normalized image $C$ can be computed by
\begin{equation}
\label{eqn1}
C=\frac{(B-B\otimes k_w)}{\sigma_w}, 
\end{equation}

where

\begin{equation}
\label{eqn2}
\sigma_w= \sqrt{ \left[(B-B\otimes k_w)^2\right]\otimes k_w}.
\end{equation}

\begin{figure}    
   \centerline{\includegraphics[width=0.95\textwidth,clip=]{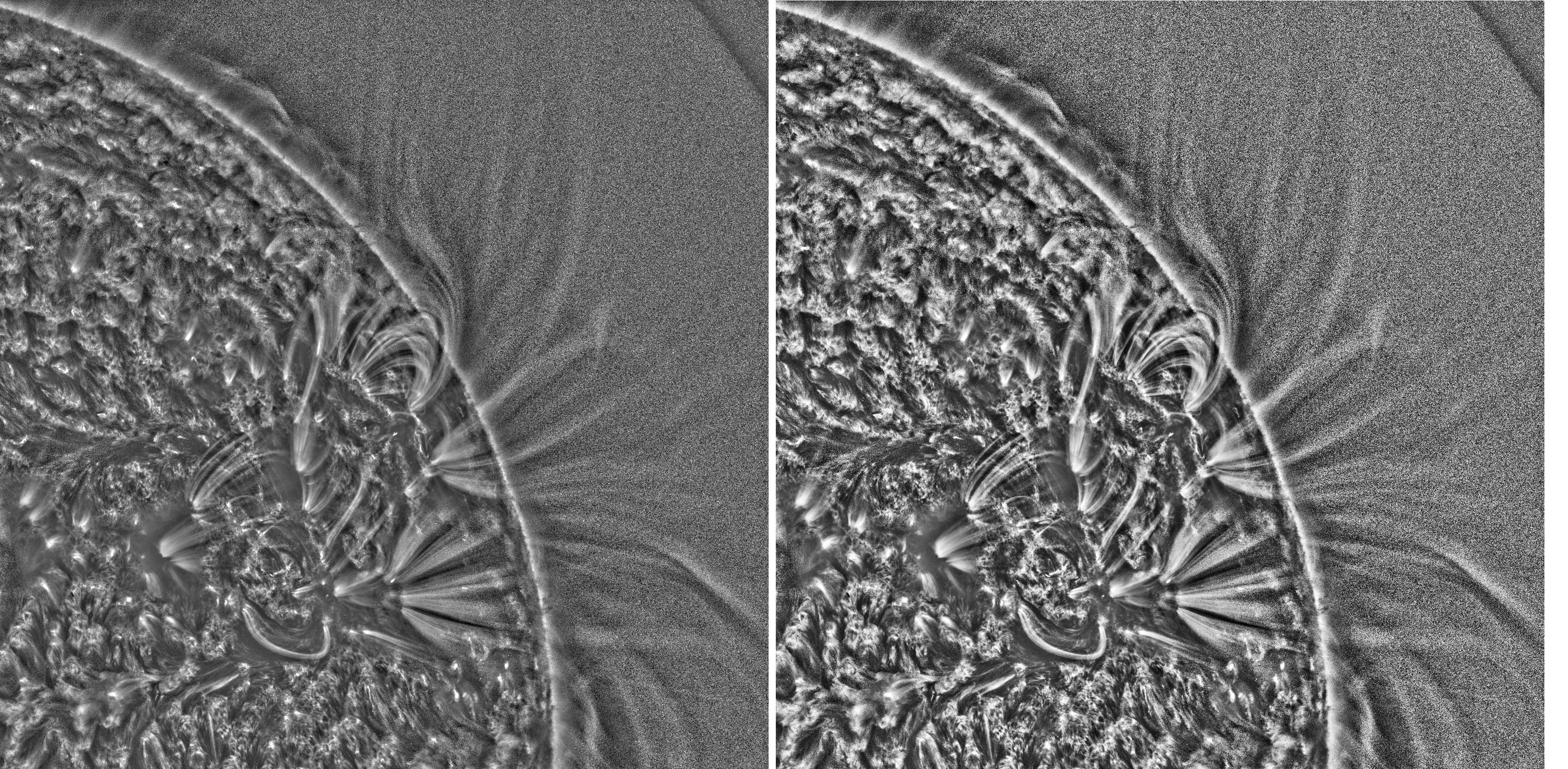}}
   \caption{Left - AIA image locally normalized by Equation (\ref{eqn1}) with $w=20$, giving image $C$. Right - the same image further processed by Equation (\ref{eqn2a}), giving image $C^\prime$.}
    \label{fig3}
  \end{figure}

For a given pixel, the numerator of Equation (\ref{eqn1}) effectively subtracts the local `mean' (weighted by the Gaussian function centered on the pixel), and the denominator $\sigma_w$ is the local `standard deviation', also weighted by the Gaussian kernel. Thus, simply speaking, $C$ is locally normalized to a standard deviation of one and a mean of zero. An example of $C$, computed with $w=20$ is shown in the left image of Figure \ref{fig3}. 

This process of local normalisation is in essence similar to adaptive histogram equalization,
where each pixel is scaled according to the statistics of the local
group of pixels. The main difference is that histogram equalization
enables precise control of the output pixel value range, whereas Gaussian normalisation does not. This problem
is solved by means of a further transformation applied to image $C$:

\begin{equation}
\label{eqn2a}
C^\prime = \arctan(k\,C).
\end{equation}

Since C is distributed across both negative and positive values, the $\arctan$
function in formula (3) serves a purpose similar to a Gamma transformation i.e.
pixel values near the zero are amplified and extreme values far from zero are
attenuated. This transform ensures the control over the output pixel value range
and prevents saturation of output pixel values. An example of $C^\prime$ is shown in the right image of Figure \ref{fig3}. The parameter $k$ controls the severity of the transformation. A default value of 0.7 gives good results for all the data shown in the results section. The pixel values of image $C$ have a standard deviation $\sim$1, and a value for $k$ of $\sim$0.7 gives sensible results.

A set of $n$ locally-normalized and $\arctan$-transformed images, $C^\prime_i$, is created for $n$ values of $w_i$. Values used throughout this work are $w=1.25, 2.5, 5, 10, 20, 40$. A global gamma-transformed image, $C^\prime_g$ is also created by 

\begin{equation}
\label{eqn2b}
C^\prime_g = \left(\frac{B-a_0}{a_1-a_0}\right)^{1/\gamma}.
\end{equation}

\noindent $\gamma$ can be set at values of 2.5 to 4. Throughout this work, we use $\gamma=3.2$, and this gives good results for all instruments and bandpasses we have processed. $a_0$ and $a_1$ set the minimum and maximum input values respectively. For processing single images, they can be set to the minimum and maximum image values. For batch processing of a large number of observation, they can be set as the minimum and maximum values of the whole set of observations. For AIA/SDO observations, these values can be easily found at the start of a batch process by reading the appropriate values stored in the data headers (that is, without having to read in the full images). 

The final processed image, $I$ is calculated by a weighted average of the $C^\prime_i$ using weights $g_i$, and addition of $C^\prime_g$ weighted by a global weight $h$:

\begin{equation}
I=hC^\prime_g + \frac{(1-h)}{n}\sum_{i=1}^{n}g_iC^\prime_i.
\label{eqn3}
\end{equation}

\begin{figure}    
   \centerline{\includegraphics[width=0.8\textwidth,clip=]{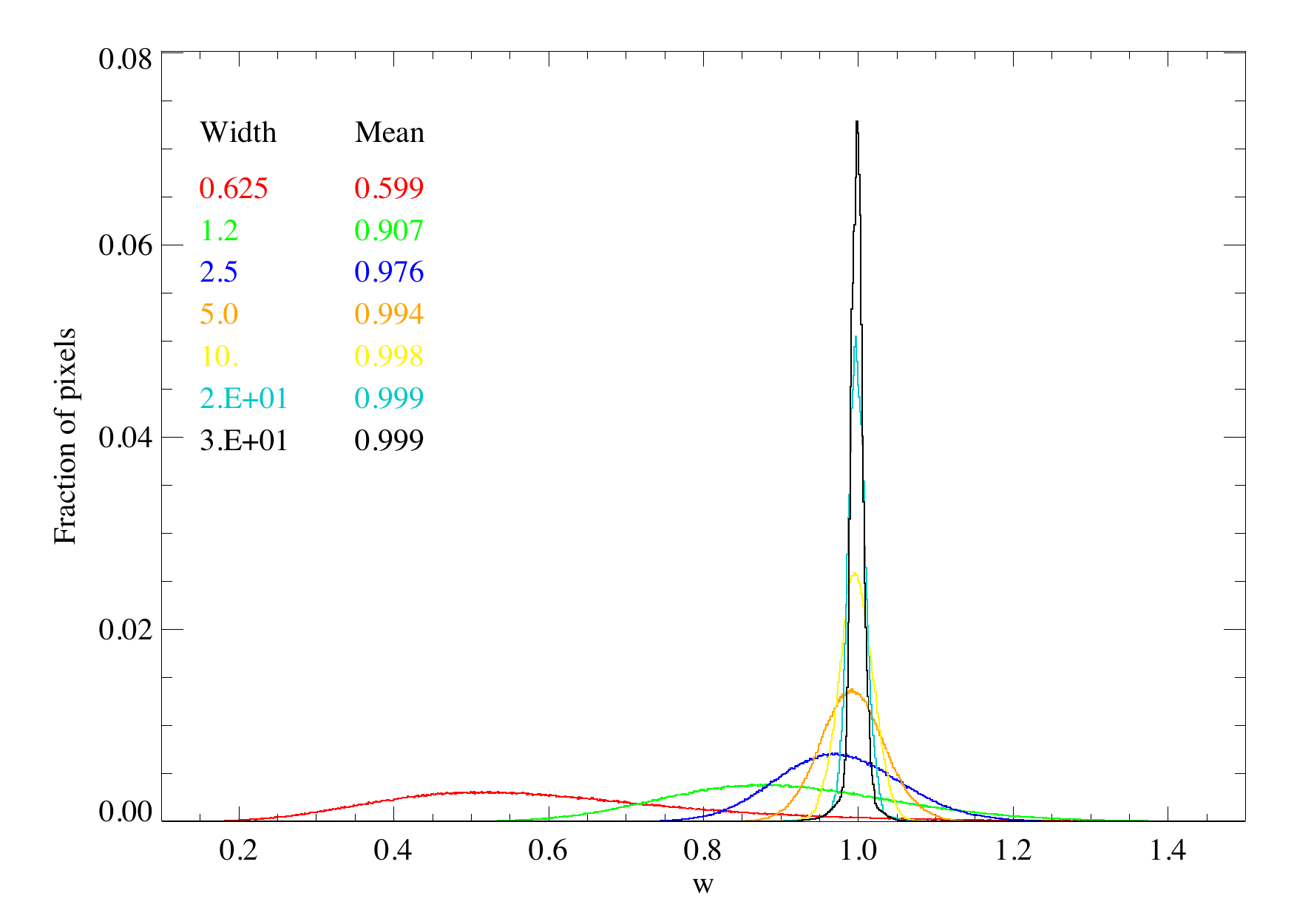}}
   \caption{The distribution of local standard deviations across an image containing only normally-distributed random noise plotted for different values of $w$.}
    \label{fig4}
  \end{figure}

The $g_i$ can be calculated by a study of images consisting of only normally-distributed random noise. For narrow Gaussian kernels, the mean of the local standard deviations across the whole image $<\sigma_w>$ is an underestimate of the global standard deviation, and there is a larger variation in the local standard deviation. This is intuitive - the ratio of the mean of the local standard deviation to the global standard deviation approaches unity with a wider kernel, and there is less variation in the local standard deviation. This is shown in Figure \ref{fig4}. Therefore the weights $g_i$ are set at smaller values for small $w$, and approach values near one as $w\gtrsim3$, as listed in Figure \ref{fig4}. As described in Equation (\ref{eqn3}), the original (normalized) image $C^\prime_g$ is included to give contextual information of the largest scale structure. We use $h=0.7$ in this work. In practice, the weights $g_i$ and $h$ may be adjusted according to the desired output, and also according to the type of input image (e.g. wavelength or channel). For most purposes, the $g_i$ can be set equal for all scales so that a straightforward mean of the locally-normalized $C^\prime_i$ is made.

\subsection{Pseudocode for MGN}

\begin{enumerate}
\item Replace spurious negative pixels with zero or local mean/median.
\item Create Gaussian kernel of width $w_i$. Kernel elements should sum to unity.
\item Convolve image with kernel to create local mean image $B\otimes k_w$.
\item Calculate difference between image and the local mean image, square the difference, and convolve with kernel. Square root the resulting image to give `local standard deviation' image $\sigma_w$ (Equation (\ref{eqn2})).
\item Calculate normalized image $C_i$ by subtracting the local mean image and dividing by the local standard deviation image (Equation (\ref{eqn1})). Store the result.
\item Apply $\arctan$ transformation on $C_i$ to give $C^\prime_i$.
\item Repeat 2-6 with the different kernel widths $w_i$.
\item Take mean, or weighted mean if preferred, of the $C^\prime_i$ to give a weighted mean locally-normalised image.
\item Calculate a global gamma-transformed image $C^\prime_g$ by applying Equation (\ref{eqn2b}).
\item Sum the weighted mean locally-normalised image with the global normalized image $C^\prime_g$, with appropriate weight $h$ (as Equation (\ref{eqn3})).
\end{enumerate}

Note that considerable efficiency is gained by creating a one-dimensional Gaussian kernel for convolving first along the $x$-direction, then convolving the resulting image along the $y$-direction with a transposed kernel. This is far more efficient than convolving directly with a two-dimensional kernel. Compared to the NAFE \cite{druckmuller2013}, or wavelet-based routines, the MGN method is a simple one, and as Gaussian smoothing is a standard practice employed by most programming languages, can readily be programmed in a few lines of code. Without great rigour, we test the efficiency of the processing using the Interactive Data Language (IDL) on a MacBook Pro laptop with 8Gb Ram and a 2.6GHz Intel Core i7 processor. The computing time increases approximately linearly from $\sim$1s for a 500$\times$500 image to 10s for a 2K$\times$2K image. A full AIA/SDO 4096$\times$4096 image takes 40s. This is extremely fast compared to other methods such as the NAFE (at least an order of magnitude faster), and does not require a high-performance computer.

  \section{Results}
  \label{results}
  
\begin{figure}    
   \centerline{\includegraphics[width=0.95\textwidth,clip=]{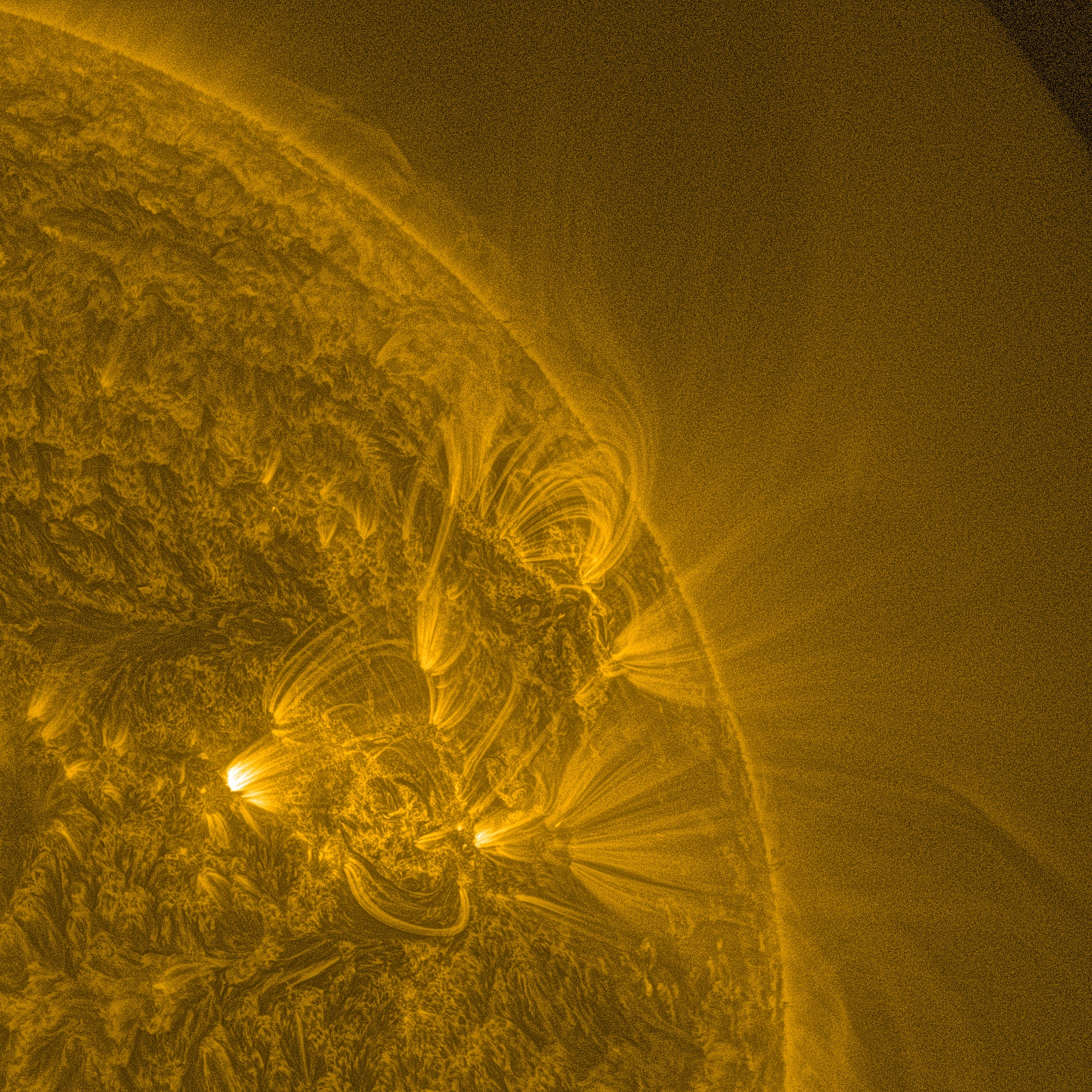}}
   \caption{AIA 171 \AA\ image taken at 04 May 2005 00:00 UT, as used as an example in the previous sections. This image has been processed with the MGN.}
    \label{fig5}
  \end{figure}

  Figure \ref{fig5} shows the result of applying MGN to the example AIA image of the previous sections. Structure is enhanced down to fine spatial scales, although large-scale context is preserved (that is, bright regions are still brighter than dark regions). Off-limb structure, where present, is enhanced without being swamped by noise. Structure is enhanced in the extremely bright region at the base of an active region (the cyan-bordered region in Figure \ref{fig2}), which originally appeared saturated. Structure can also be seen in dark regions immediately neighbouring the bright region. It is also very useful to be able to trace structure from origins on the disk to off-limb heights. In this example, there are loops and fan-like bright features which extend from the disk and past the limb. Movies created using the processing reveal dynamic features at very small scales such as flows within filament channels, flows along large coronal loops, small-scale activity within active regions, and movement in extended off-limb structures in response to dynamic events. 
  
An example AIA movie is available online (aia\_movie.gif - an animated gif which can be viewed with a web browser) which shows a large portion of the west disk and off-limb region observed in the 171 \AA\ channel during 18 January 2013. Activity and small-scale flows are clearly seen, alongside larger-scale eruptions, and the movement/expansion of loops off the limb out to the edge of the field of view.
  
Another online movie (comparison\_movie.gif) compares the appearance of a sequence of AIA observations processed by the MGN (left panel), and by a simple gamma transformation (right panel). This sequence shows a flaring active region and the eruption of a CME observed in the 171 \AA\ channel during 08 March 2011 in the South-West. The disk and off-limb structure is seen far more clearly in the MGN images. Individual loops are resolved, even in the complicated off-limb region to the North of the active region. In the original data, the sequence is dominated by the bright core of the active region, and when the flare peaks, by the small saturated regions centered on the flare. The MGN images show this activity clearly, and also succeed in maintaining a clear and stable view of surrounding structure throughout the movie, including the `wobble' of the active region loops in response to flaring activity. This enables a more comprehensive analysis of the event, showing connections between activity and movement of structure which can be hidden without MGN processing.
  
\begin{figure}    
   \centerline{\includegraphics[width=0.9\textwidth,clip=]{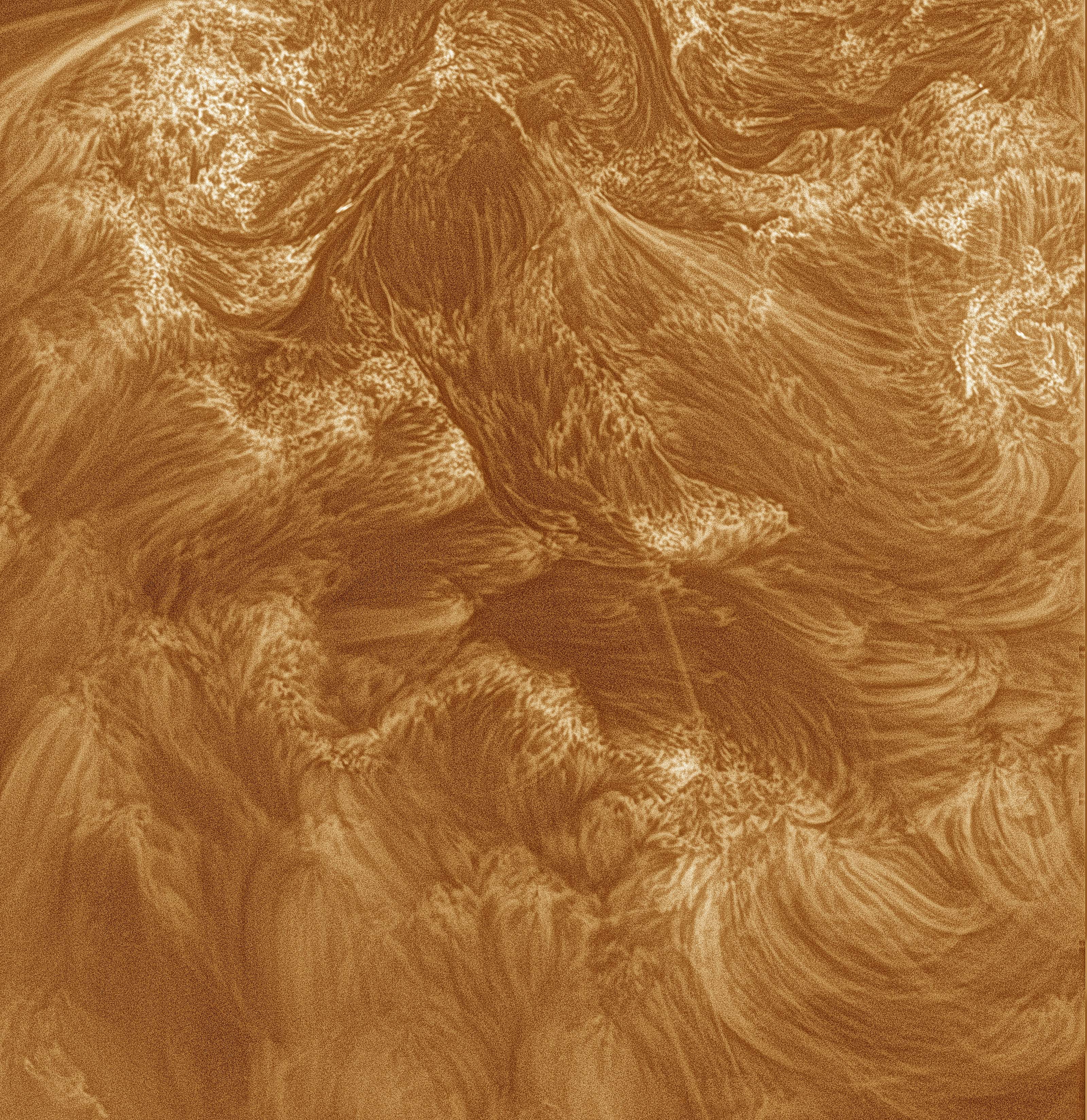}}
   \caption{HIC image of a region on the disk in 193 \AA. Hi-C was flown on a rocket during 11 July 2012, and gained several minutes of observation of this shown field of view. It has a 0.1" spatial resolution, and a cadence of $\sim$5s.}
    \label{fig6}
  \end{figure}

\begin{figure}    
   \centerline{\includegraphics[width=0.9\textwidth,clip=]{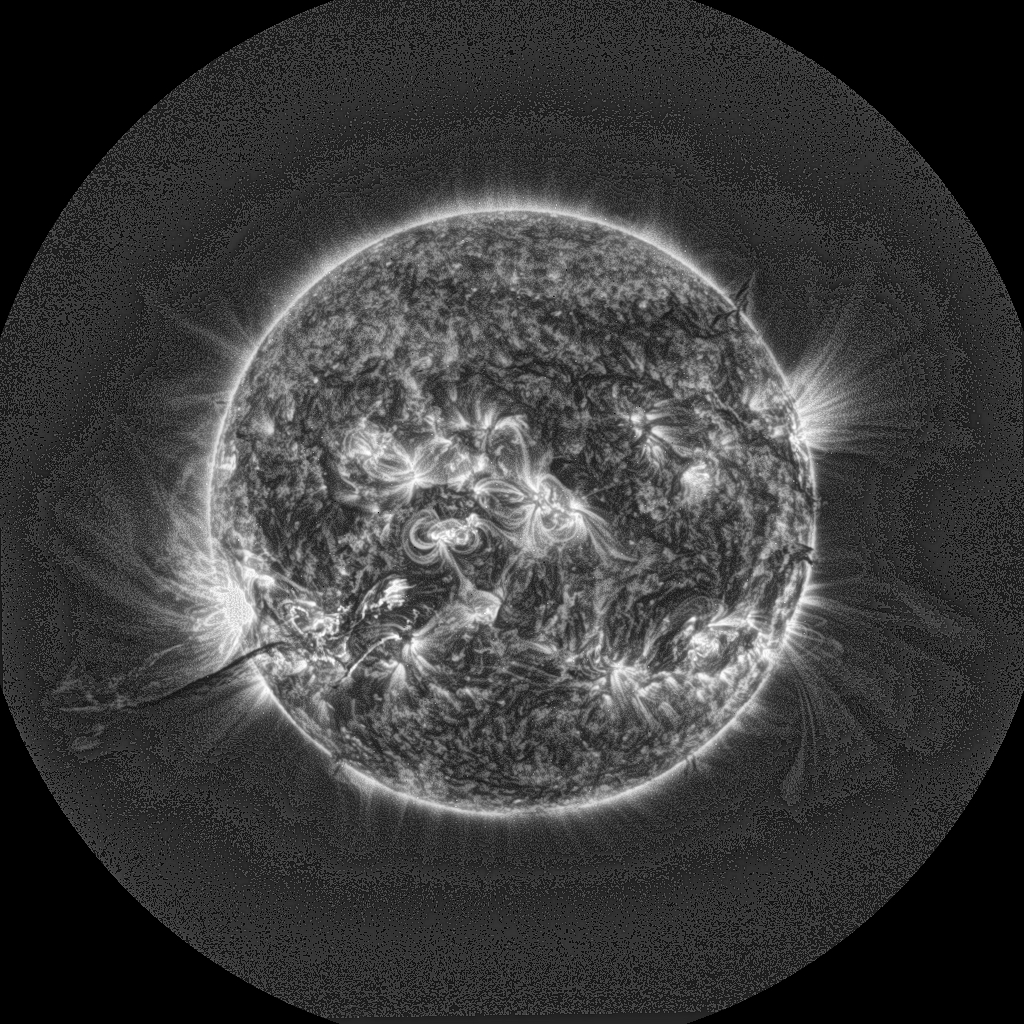}}
   \caption{MGN-processed SWAP image of the whole disk and off-limb region taken during 31 August 2012.}
    \label{fig7}
  \end{figure}

  Figure \ref{fig6} shows a Hi-C image processed with MGN. A movie showing the northernmost portion of this image is available online (hic\_movie.gif). Various features such as magnetic braiding \cite{cirtain2013} are clearly seen, as are plasma streams along the dark filament channels, and small-scale movement within the magnetic `moss'.  Such features cannot be seen without appropriate processing. Figure \ref{fig7} shows the application of MGN to an observation by the \textit{Sun Watcher using APS} (SWAP) instrument aboard the \textit{PRoject for OnBoard Autonomy 2} (PROBA2) satellite \cite{berghmans2006,seaton2013}. This instrument has a coarser resolution than AIA, but has an extended field of view. Figure \ref{fig7} reveals an erupting filament out to the extremity of the field of view, and other quiescent structures to $\sim$1.5\Rs. Again, even low-signal structures are enhanced without too much amplification of noise.
  
\begin{figure}    
   \centerline{\includegraphics[width=1.0\textwidth,clip=]{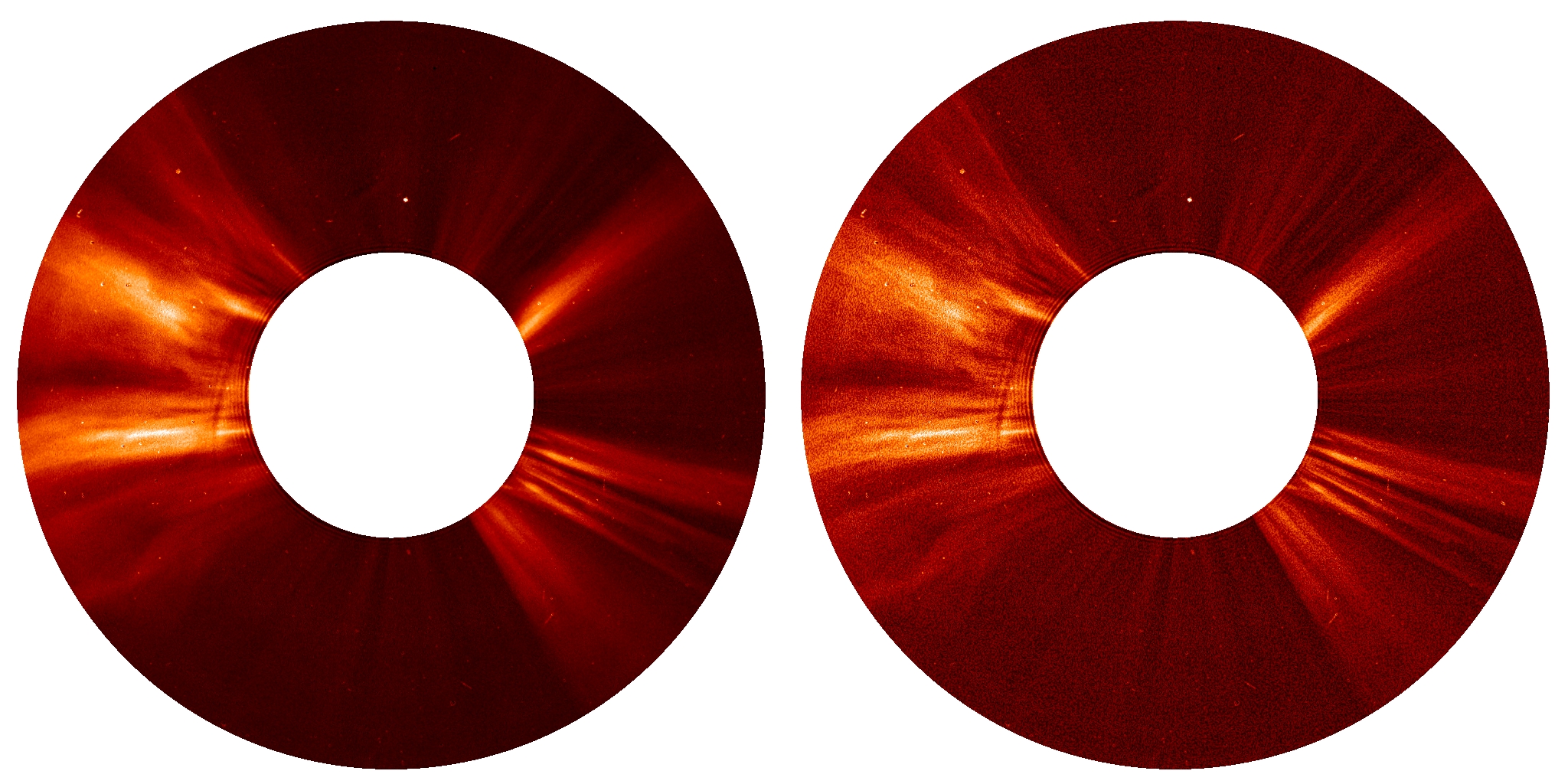}}
   \caption{Left - LASCO C2 image taken during 14 January 2011. The field of view extends from 2.2 to 5.9\Rs. Right - the same image processed with the MGN.}
    \label{fig8}
  \end{figure}

The MGN processing is not limited to EUV observations. Figure \ref{fig8} shows its application to a white light coronagraph image of the extended inner corona by the \textit{Large-Angle and Spectrometric Coronagraph} (LASCO) aboard SOHO. The left image shows an observation of 14 January 2011, with a long-term background subtracted to remove instrumental stray light and F-corona. A point filter has also been applied to reduce spurious bright pixels. The right image shows a MGN-processed image, with parameters $k=0.8, h=0.9$ and $\gamma=1$. Smaller-scale features are enhanced, in particular faint plumes over the poles which are difficult to see in the original image. In this respect, the MGN is an improvement over the NRGF \cite{morgan2006}, although the NRGF provides structural context closer to the true K-corona.

It is difficult to make a qualitative comparison of images created using different processing. Different processes may serve different analysis purposes - the important function for the process described here is to reveal information hidden in the original images. A non-rigorous comparison `by eye' of the same images processed using the MGN and of the NAFE \cite{druckmuller2013} suggests that the results are similar, with the NAFE giving slightly better clarity overall, and improved noise suppression - most obvious at large heights off the limb. For the MGN, noise reduction arises naturally from taking the (weighted) mean across many spatial scales. The noise suppression is not as effective as the NAFE, and the MGN method lacks control over the degree of noise suppression. The NAFE method is also better suited to reveal contrast in very bright regions.

\section{Summary}
\label{summary}
The MGN method normalizes an image by using the local mean and standard deviation calculated using a Gaussian-weighted sample of local pixels. This normalised image is transformed by the $\arctan$ function (similar to a Gamma transformation). This is applied over several spatial scales, and the final image is a weighted combination of the normalized components. The results compare well with multiresolution wavelet enhancement or the NAFE procedure, but is far more computationally efficient. We hope that the MGN will become an established tool for researchers, offering a good compromise between computational time and clarity of the final images. The method is simple to implement, and the author is happy to provide the IDL code by email request.

\begin{acks}
We are grateful for comments by the anonymous referee which improved this work. Huw Morgan is grateful for funding from the Coleg Cymraeg Cenedlaethol to Prifysgol Aberystwyth and support of SHINE grant 0962716 and NASA grant NNX08AJ07G to the Institute for Astronomy, University of Hawaii. The work of Miloslav Druckm\"uller was supported by Grant
Agency of Brno University of Technology, project FSI-S-11-3.  We acknowledge the High resolution Coronal Imager instrument team for
making the flight data publicly available. MSFC/NASA led the mission
and partners include the Smithsonian Astrophysical Observatory in
Cambridge, Mass.; Lockheed Martin's Solar Astrophysical Laboratory in
Palo Alto, Calif.; the University of Central Lancashire in Lancashire,
England; and the Lebedev Physical Institute of the Russian Academy of
Sciences in Moscow.
\end{acks}
  
\bibliographystyle{spr-mp-sola}


\end{article} 

\end{document}